%% file: 1201main.tex
\documentclass{aa}
\usepackage{psfig}
\usepackage{natbib}
\usepackage{txfonts}

\begin{document}
\title{Relationship between X-ray and ultraviolet emission of flares
  from dMe stars observed by \textit{\textbf{XMM-Newton}}} 
\author{U.~Mitra-Kraev\inst{1} \and L.~K.~Harra\inst{1} \and
  M.~G\"udel\inst{2} \and M.~Audard\inst{3} \and
  G.~Branduardi-Raymont\inst{1} \and H.~R.~M.~Kay\inst{1} \and
  R.~Mewe\inst{4} \and A.~J.~J.~Raassen\inst{4,5} \and L.~van
  Driel-Gesztelyi\inst{1,6,7}}  
\offprints{U.~Mitra-Kraev,\\\email{umk@mssl.ucl.ac.uk}}
\institute{
Mullard Space Science Laboratory, University College London, Holmbury
  St.~Mary, Dorking, Surrey RH5 6NT, UK  
\and 
Paul Scherrer Institut, W\"urenlingen \& Villigen, 5232 Villigen PSI,
  Switzerland  
\and 
Columbia Astrophysics Laboratory, Columbia University, 550 West 120th
  Street, New York, NY 10027, USA  
\and 
SRON National Institute for Space Research, Sorbonnelaan 2, 3584 CA
  Utrecht, The Netherlands  
\and Astronomical Institute `Anton Pannekoek', Kruislaan 403, 1098 SJ
  Amsterdam, The Netherlands  
\and 
Observatoire de Paris, LESIA, 92195 Meudon, France 
\and 
Konkoly Observatory, 1525 Budapest, Hungary
} 
\date{Received 30 April 2004 / Accepted 30 September 2004} 
\titlerunning{Relationship between X-ray and UV emission of stellar flares}
\authorrunning{U.~Mitra-Kraev et al.}
\abstract{
\input{1201abst}
} 
\maketitle

\section{Introduction} 
\label{intro}
\input{1201intr}

\section{Observations} 
\label{obs}
\input{1201obs}

\section{Results} 
\label{res}
\input{1201res}

\section{Discussion} 
\label{disc}
\input{1201disc}

\section{Conclusions} 
\label{conc}
\input{1201conc}

\begin{acknowledgements}
We kindly thank the anonymous referee for valuable comments and suggestions. 
UMK, LKH and HRMK acknowledge financial support from the UK Particle
Physics and Astronomy Research Council (PPARC).  
General stellar X-ray astronomy research at PSI has been supported by
the Swiss National Science Foundation under project 20-66875.01.  
MA acknowledges support from NASA to Columbia University for
\textit{XMM-Newton} mission support and data analysis.  
LvDG acknowledges the Hungarian government grant OTKA T-038013. 
\\ \\
In memory of Rolf Mewe, who sadly and unexpectedly passed away after
this paper was submitted. 
\end{acknowledgements}

\bibliographystyle{aa}
\bibliography{1201ref}

\end{document}

%% file: 1201abst.tex
We present simultaneous ultraviolet and X-ray observations of the
dMe-type flaring stars \object{AT Mic}, \object{AU Mic}, \object{EV
Lac}, \object{UV Cet} and \object{YZ CMi} obtained with the {\it
XMM-Newton} observatory.  
During 40 hours of simultaneous observation we identify 13 flares
which occurred in both wave bands.  
For the first time, a correlation between X-ray and ultraviolet flux for
stellar flares has been observed. 
We find power-law relationships between these two wavelength bands for
the flare luminosity increase, as well as for flare energies, with
power-law exponents between 1 and 2. 
We also observe a correlation between the ultraviolet flare energy and
the X-ray luminosity increase, which is in agreement with the Neupert
effect and demonstrates that chromospheric evaporation is taking place.


%% file: 1201intr.tex
{\sl XMM-Newton} is an ideal platform for observing cosmic sources
simultaneously in X-rays and ultraviolet (UV).  
In particular, flux and energy comparisons between these wavebands are
important in understanding the flare mechanisms in stellar
coronae. Late-type stars typically show high coronal activity similar
to the Sun, with flux variability through all observed wavelength
bands.  
dMe-type stars are particularly active. 
They show frequent flaring activity \citep{pallavicini1990}, as well
as strong emission lines (including the H$\alpha$ Balmer emission
line, which is denoted by the `e' in dMe).  
Soft X-rays ($<$12 keV) typically originate in the corona, while UV
radiation comes from the chromosphere and transition region.  
Flares are observed in both wave bands, but what is the physical
process that connects X-ray and UV flares?  

In the chromospheric evaporation picture \citep[see
e.g.][]{antonucci1984}, flares are believed to be due to  the energy
release from the reconnection of magnetic field lines in the lower
corona.  
Electrons are accelerated at the reconnection site. 
They gyrate downward along the magnetic field lines, emitting
gyrosynchrotron radio emission.  
Their collision with denser material in the chromosphere unleashes
bremsstrahlung seen in hard  X-rays ($>$20~keV).  
Both the gyrosynchrotron radio and hard X-ray flare emissions are
impulsive, with a fast increase and a steep decay.  
At the same time, the electrons impulsively heat the chromosphere,
which results in prompt optical/UV emission which closely correlates
with the hard X-ray emission \citep{hudson1992}.  
During the collisions, the chromospheric ions get further ionised, the
material heats up and evaporates, increasing the density and the
temperature of the reconnected loops in the corona.  
The hot material is seen in soft X-ray and extreme ultraviolet
emissions, where the light curves are more gradual, with a slower
increase than the impulsive emission and a much longer, exponential
decay.  
The impulsive and gradual emissions are often temporally connected
through the``Neupert effect'' relation \citep{neupert1968,dennis1993}  
\begin{equation} \label{eq_neupert}
L_{\rm grad} (t) = q \cdot \int_0^t L_{\rm impuls}(t')\hspace{1ex} dt',
\end{equation}
indicating that the gradual radiative loss rate is directly
proportional to the cumulative impulsive energy input, and suggesting
that the energy input from the non-thermal electrons is responsible
for the heating of the plasma.   
\input{1201tab1}

The soft X-rays originating from the hot ($>$1~MK) coronae of
late-type stars are dominated by emission lines (\ion{H}- and
\ion{He}-like transitions of \ion{C}, \ion{N}, \ion{O}, \ion{Ne},
\ion{Mg}, and \ion{Si}, and \ion{Fe} K- and L-shell transitions),
indicating thermal processes. 
The UV emission from dMe stars is mainly line emission formed around
$10^4$~K \citep{linsky1982}, though during flares continuum emission
is also observed in the near-optical UV band \citep{hawley1991}.  
At optical wavelengths, both impulsive continuum and gradual line
emission are observed during flares \citep{garcia-alvarez2002}.  
The Neupert effect has been known to exist between impulsive U-band
and soft X-ray emission \citep{guedel2002a,hawley1995}.  

Previous investigations relating UV and soft X-rays in late-type stars
were either focused on individual flares or statistical flux-flux
relationships for the entire stellar emission with no temporal
resolution.  
The UV data were collected with the International Ultraviolet Explorer
({\sl IUE}), which obtained low dispersion spectra in the
1150--3200~\AA ~band, and compared with data from the X-ray satellites
{\sl Einstein} (0.2--4~keV), {\sl EXOSAT} (0.06--2~keV) and {\sl ROSAT}
(0.1--2.4~keV), and more recently UV data from the Hubble Space
Telescope were compared with X-ray data from the Chandra satellite
\citep{ayres2001}.  
To distinguish flaring from non-flaring conditions,
\citet{mathioudakis1989} compared flux-flux relationships of inactive
dM/dK to active dMe/dKe stars, using the latter as a proxy for flaring
conditions.  
They find a power-law relationship with a slope of $\approx$1 between
chromospheric \ion{Mg}{ii}~h~\&~k (2795--2803 \AA) {\sl IUE} and
coronal X-ray ({\sl Einstein} and {\sl EXOSAT}) flux, but only for the
active stars.   
The inactive stars are scattered, with generally lower X-ray luminosity. 
Thus, the coronal X-ray emission is enhanced for stars with flares
compared to stars without. 
Similar analyses for F-K type stars have been made by \citet[][
including a basal flux subtraction]{schrijver1992} and \citet[][ fluxes
scaled to bolometric fluxes]{ayres1995}, who found power-law slopes
between 1.5 and 2.9. 
None of these papers tested the relationship of different fluxes for
individual flares.  

We set out to probe directly the statistical flux-flux relationship
between X-ray and UV emission for flares.  
dMe-type stars are the natural choice, as they flare frequently.
The instruments on-board {\sl XMM-Newton} provide an excellent
opportunity for carrying out these observations.  

This paper is structured in the following way: Sect.~\ref{obs}
describes the observations and the instrumental setup.  
The results are presented in Sect.~\ref{res}, starting with the X-ray
and UV luminosity light curves of \object{EV Lac}, \object{UV Cet},
\object{YZ CMi}, \object{AU Mic} and \object{AT Mic}.   
The respective light curves are cross-correlated and flares identified
in both wavebands.  
Then, the power-law relationships between luminosity increase per
flare and flare energy are presented.  
We discuss the results in Sect.~\ref{disc} and conclude with
Sect.~\ref{conc}.  


%% file: 1201tab1.tex
\begin{table*} [htb] \caption{Observational Parameters and
      Results} \label{targets} 
\begin{minipage}{10cm}
\renewcommand{\thefootnote}{\thempfootnote}
\begin{tabular}{cccccc}
\hline 
\hline 

& \object{EV Lac} & \object{UV Cet} & \object{YZ CMi} & \object{AU Mic} &
\object{AT Mic}  \\  
\hline 

Index & A & B & C & D & E \\

Spectral Type & dM4.5e & dM5.5e+dM5.5e & dM4.5e & dM1e & dM4.5e+dM4.5e
\\
Distance $d_\star$ [pc] & 5.0\footnote{\cite{perryman1997}} &
2.6\footnote{\cite{harrington1980}}\addtocounter{mpfootnote}{-1} & 
5.9\footnotemark[1] & 9.9\footnotemark[1] & 10.2\footnotemark[1] \\ 

UV Filter & UVW1 & UVW1 & UVW2 & UVW2 & UVW2 \\
UV Range [\AA] & 2450--3200 & 2450--3200 & 1800--2250 & 1800--2250 &
1800--2250  \\ 

\hline

\rule{0mm}{4mm}$\bar{L}_{\rm uv} [10^{29}{\rm erg/s}]$ & 0.434 &
0.0258 & 1.28 & 12.3 & 6.39 \\  

$\bar{L}_{\rm x} [10^{29}{\rm erg/s}]$ & 0.545 & 0.0333 & 0.370 & 2.99
& 2.89 \\   

Time-lag UV$-$X-ray [s] & $-600_{-600}^{+600}$ &
$-200_{-200}^{+600}$ & $-600_{-400}^{+600}$ &
$-400_{-400}^{+400}$ & $-1000_{-1000}^{+800}$ \\ 

X-ray/UV max.~Correlation & 0.68 & 0.90 & 0.78 & 0.68 & 0.84 \\
\hline
\end{tabular} 
\end{minipage}
\end{table*}

%% file: 1201obs.tex
We present {\it XMM-Newton} data of dMe-type stars obtained during the
Reflection Grating Spectrometer (RGS) Guaranteed Time Programme. 
We make use of the pn-European Photon Imaging Camera 
\cite[EPIC-pn,][]{strueder2001}, and the Optical Monitor
\citep[OM,][]{mason2001}.  
The EPIC-pn covers the range 0.2--12~keV (1--62~\AA).  
The OM observed with two different UV filters, one for each
observation, namely the UVW1 (2450--3200~\AA) and the UVW2 (1800--2250~\AA)
filter. 

Table \ref{targets} gives a list of the targets, shows the spectral
type, the distance to the star (from SIMBAD parallaxes), the observed
UV range and UV filter.  
The index is for further reference in the later plots. 
While some of the X-ray results of these targets have already been
published \citep{raassen2003,ness2003,magee2003}, this is the first
time that we make use of the simultaneously obtained UV data.  

For all observations, OM observed in IMAGE mode, where a single
exposure lasted 800~s and the dead-time between two exposures was
320~s, thus resulting in a time resolution of 1120~s.  
The data for the EPIC-pn light curves were binned into 200~s time
intervals.   
All data were reduced with the {\sl XMM-Newton} Science Analysis
Software (SAS) version~5.4.  


%% file: 1201res.tex
\subsection{Light curves}
\subsubsection{Determining the luminosity} \label{3.1.1}
\input{1201fig1}
\input{1201fig2}
Figures \ref{figure1a} and \ref{figure1b} (left panels) display the
X-ray and UV light curves of \object{EV Lac}, \object{UV Cet},
\object{YZ CMi}, \object{AU Mic} and \object{AT Mic}.
In the ultraviolet, \object{EV Lac} and \object{UV Cet}
(Fig.~\ref{figure1a}) were observed with the UVW1 filter, whereas
\object{YZ CMi}, \object{AU Mic} and \object{AT Mic}
(Fig.~\ref{figure1b}) were observed with the UVW2 filter.  

The conversion from X-ray and UV count rates to luminosity has been
made in the following way: 
The UV luminosity ${L}_{\rm uv}$ is given by
\begin{equation}
  {L}_{\rm uv} =  4 \pi d_\star^2 \cdot G \cdot 
  \Delta_{\rm uv} \cdot c_{\rm uv},
  \label{eq1} 
\end{equation} 
with $c_{\rm uv}$ the UV count rate and $d_\star$ the distance to the
star (see Table~\ref{targets}).
$G$ is the factor to convert from count rate to flux, derived by
folding stellar spectra with the in-flight response curves of the
OM and tabulated on the {\sl XMM-Newton} SAS homepage at Vilspa. 
The values of $G$ for M-type stars are $7.32(\pm 0.39)\cdot 10^{-16}~{\rm
  erg/cm^2/\AA/count}$ for the UVW1 filter and $1.04(\pm 0.21)\cdot
10^{-13}~{\rm erg/cm^2/\AA/count}$ for the UVW2 filter. 
$\Delta_{\rm uv}$ is the bandwidth of the UV filter, 750~\AA~for UVW1 and
450~\AA~for UVW2.  
The values of the average stellar UV luminosities over the whole
observations, $\bar{L}_{\rm uv}$, are displayed in Table~\ref{targets}.

For the X-ray luminosity, the average luminosity $\bar{L}_{\rm x}$
has been obtained from fitting the EPIC-pn spectrum from the whole
observation with a 3-Temperature CIE (collisional ionisation equilibrium)
model within the SPEX package \citep{kaastra2002}
with variable elemental abundances.  
The precise values of the abundances do not matter in this context, as 
the value for the average luminosity, which is directly
given by the count rate in the spectrum, is stable.
The value of $\bar{L}_{\rm x}$ for each target, which includes dead-time
corrections, is displayed in Table
\ref{targets}.    
The  three temperatures are typically 2--3~MK, 7--8~MK and $>$20~MK. 
The X-ray luminosity $L_{\rm x}$ is then given by the average luminosity
$\bar{L}_{\rm x}$ divided by the average count rate $\bar{c}_{\rm x}$
times the X-ray count rate $c_{\rm x}$ 
\begin{equation}
  {L}_{\rm x} = \frac{\bar{L}_{\rm x}}{\bar{c}_{\rm x}} \cdot c_{\rm
  x}. \label{eq2} 
\end{equation}
One might argue that the different shapes of the quiescent and flaring
X-ray spectrum have a non-linear effect on the scaling of the
luminosity.   
The X-ray spectrum is dominated by lines in the range of 0.2--1~keV and has a
decreasing exponential tail above 1~keV.  
During flares, the exponential tail falls off more slowly than during
quiet times, indicating higher temperatures and energies, while the
low-energy range rises uniformly.
Our fits show that most of the energy comes from the low-energy range. 
The values for the total observations are 0.2--1~keV $\sim$ 70$-$75\%,
1--2~keV $\sim$ 15$-$20\%, and 2--12~keV $\sim$ 10\%, while during
flares the contribution to the lowest energy band is reduced by about
5\%, and increased by about 5\% for the highest energy band. 
This suggests that the approximation that the luminosity scales
linearly with the count rate is a fair assumption.

\subsubsection{Cross-correlation between the UV and X-ray flux}
\input{1201fig3} 
The two light curves from each of the five observations show a good
correlation. 
For an increase in the UV, we generally also see an increase in
the X-rays.
Each pair of light curves has been cross-correlated in such a way that
the 200s-binned X-ray light curve (lowest panels in the light curve
plots of Figs.~\ref{figure1a} and \ref{figure1b}) was shifted by a
multiple of 200~s, then binned to the UV resolution and correlated
with the UV light curve (upmost panel). 
The plots on the right hand side of Figs.~\ref{figure1a} and
\ref{figure1b} show the cross-correlation for each target. 
The abscissa displays the time-lag between the UV and X-ray light
curves, the ordinate the corresponding correlation coefficient. 
For all observations, the function has a distinct peak with a maximum
correlation coefficient between 0.68 and 0.90. 
For the error estimate in time, the width of the cross-correlation
distribution function at the $90\%$ level of the correlation coefficient
maximum was calculated (horizontal dotted lines in the cross-correlation
plot). This error corresponds to about the $1\sigma$ error of the
Fisher z-test. 
In all but the UV Cet observation, which is dominated by short X-ray
flares which are below the OM resolution, the entire peak within the
error lies left of zero, indicating that flares tend to peak earlier
in UV than in X-rays.  
The middle panels of the light curve plots show the X-ray light curves
at maximum correlation (shifted by the time-lag) and binned to UV
resolution.

\subsubsection{Flare identification} \label{fid}
In both the UV and the X-ray (low 800s-resolution) light curves (upper
two panels in Figs.~\ref{figure1a} and \ref{figure1b}) flares are
identified. The flare criteria are: 
(i) It has to appear in both wave bands simultaneously. 
(ii) The difference between flare peak and start intensity exceeds
$3\sigma$ of the noise (3$\times$average noise). 
(iii) The difference between flare peak and end intensity exceeds
$3\sigma$ in at least one of the light curves (and $1\sigma$ in the
other).
The beginnings and ends of the flares are marked with a dashed
vertical line, the flare peak with a dotted vertical line. 
Each flare is named by a letter and a number displayed under the UV
light curve.

\subsection{The UV--X-ray relationship in flares} \label{ffr}
For each flare, the background-subtracted energy from flare start to 
end $E_f = \int_{f_{\rm start}}^{f_{\rm end}} \left( L(t)-L_f^{\rm
    min} \right) dt$ as well as the luminosity increase from flare
start to flare peak have been obtained for both UV and X-rays (low
800s-resolution).  
Figure~\ref{regress_uvw1} shows the relationship between UV and X-rays
for the observed flares from \object{EV Lac} and \object{UV Cet} (UVW1
filter, left side) and from \object{YZ CMi}, \object{AU Mic} and
\object{AT Mic} (UVW2 filter, right side).  
The upper panels are for the luminosity, the lower panels for
the energy relationship. 
We define the spectral luminosity density (luminosity per unit
wavelength) $\mathcal{L}:={L}/{\Delta}$ and the spectral energy
density $\mathcal{E}:={E}/{\Delta}$, with $\Delta$ being the width of
the respective passband. 
The flare data are fitted with a power-law
\begin{equation}
\mbox{\rm (X-ray)} = 10^c \cdot{\rm (UV)}^\kappa,
\end{equation}
the values of $c$ and $\kappa$ are given in Table~\ref{fit}. 
\input{1201tab2}
The regressions have been made in the following way: 
For each regression, we have first applied two linear ordinary
least-square (OLS) fits \citep{isobe1990} to the logarithmic data, one
for Y dependent on X, taking the Y-error into account, the other for X
dependent on Y, taking the X-error into account (linfit.pro in
IDL). Given the two regressions (slopes $\beta_1$ and $\beta_2$), the
bisector was calculated (slope $\beta_3 = {\rm tan}\frac{{\rm
    atan}\beta_1+{\rm atan}\beta_2}{2}$) and the new variances
obtained from Monte-Carlo simulations. The OLS bisector is thus chosen
as the optimal regression.

For the luminosity (Eqs.~(\ref{eq1}) and (\ref{eq2})), we only
considered the error from the counts, which basically follows
Poisson statistics.  
Three further errors of the quantities used to derive the luminosities
do not contribute much to the overall result.
The uncertainty in stellar distance ($<5\%$) affects UV and X-ray
luminosity alike and therefore does not influence the power-law slope.  
The $G$-factor of Eq.~(\ref{eq1}) carries an error of $5\%$ for UVW1
and $20\%$ for UVW2.
As our targets all have similar spectral types, the possible systematic error
is unlikely to affect the power-law exponent either.
The third error is the (small) non-linear contribution from the
determination of the X-ray luminosity, which has been neglected (see
Sect.~\ref{3.1.1}).


%% file: 1201fig1.tex
\begin{figure*}
\centering
\vbox{
\hbox{
\psfig{figure=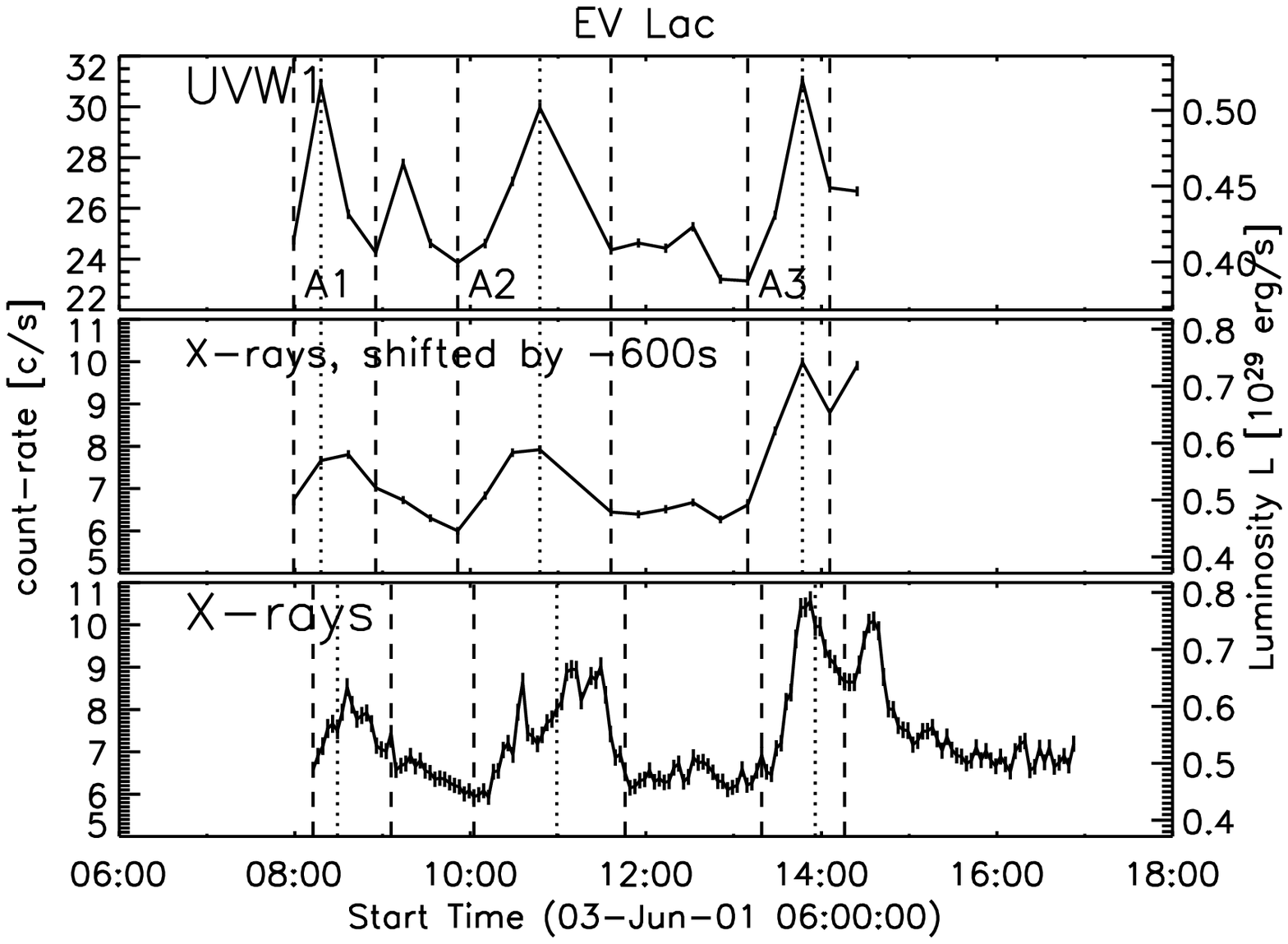,height=6.0cm,width=10.0cm}
\hspace{0.5cm}
\psfig{figure=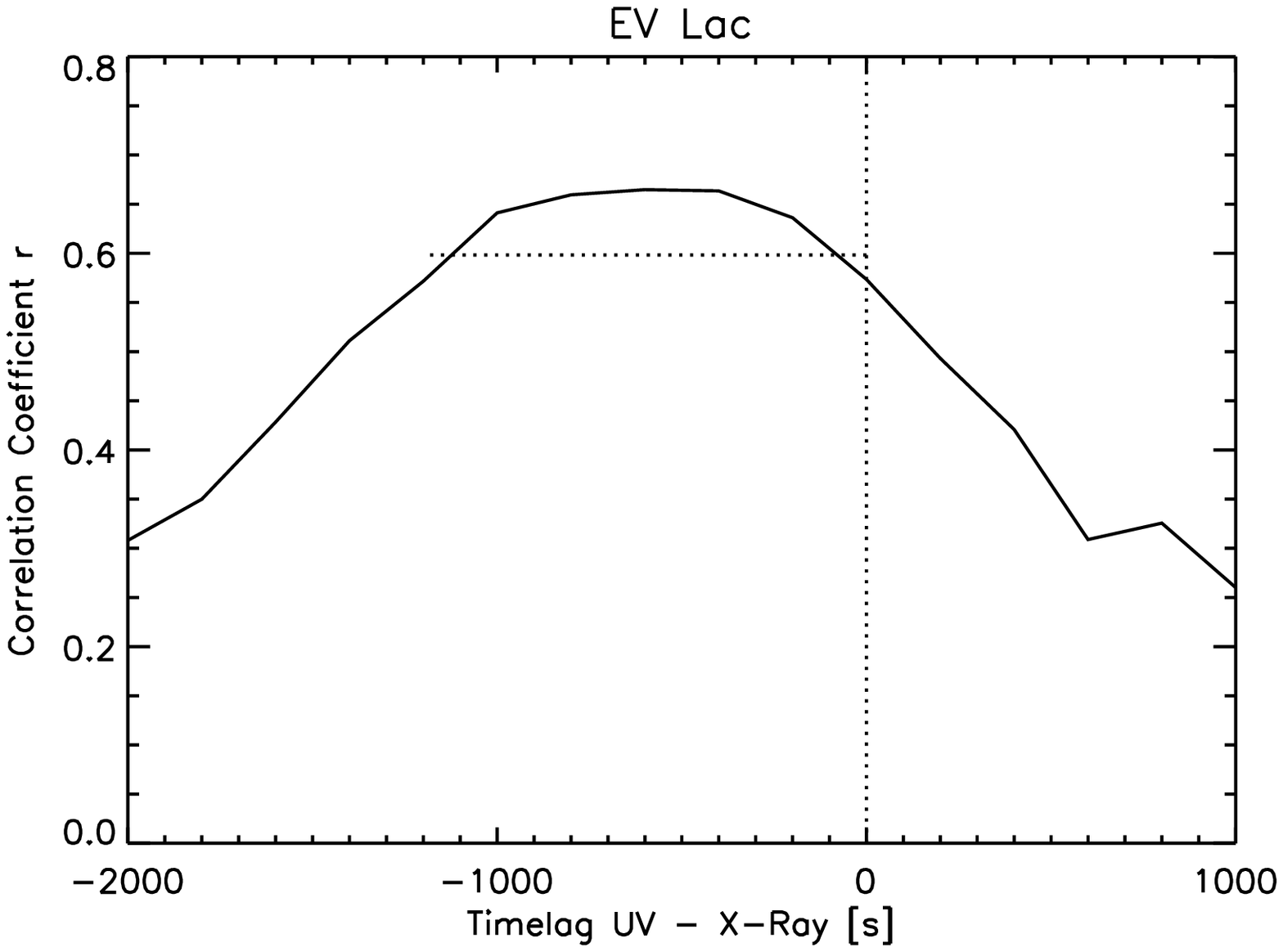,height=6.0cm,width=7.0cm}
}
\vspace{0.5cm}
\hbox{
\psfig{figure=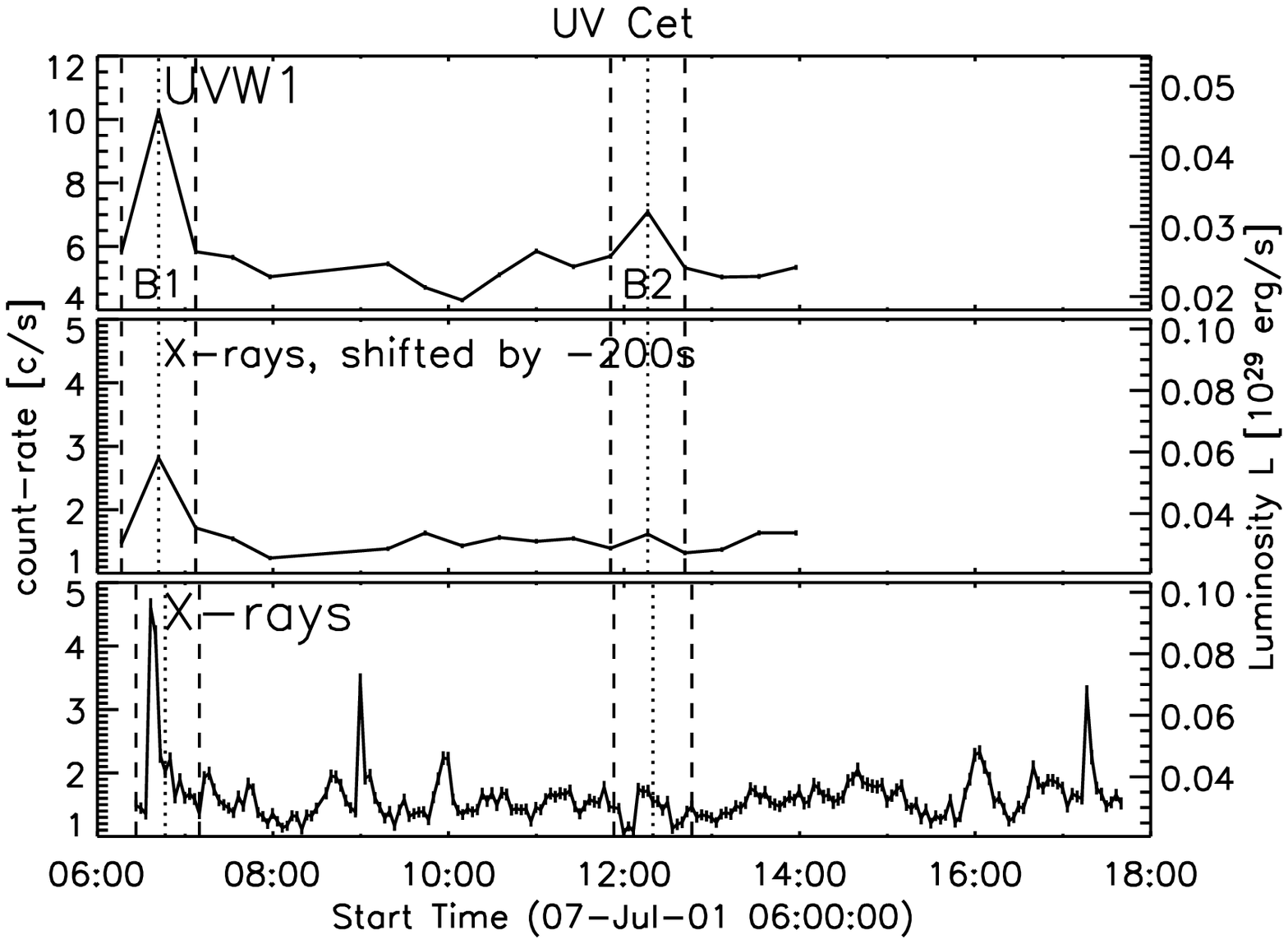,height=6.0cm,width=10.0cm}
\hspace{0.5cm}
\psfig{figure=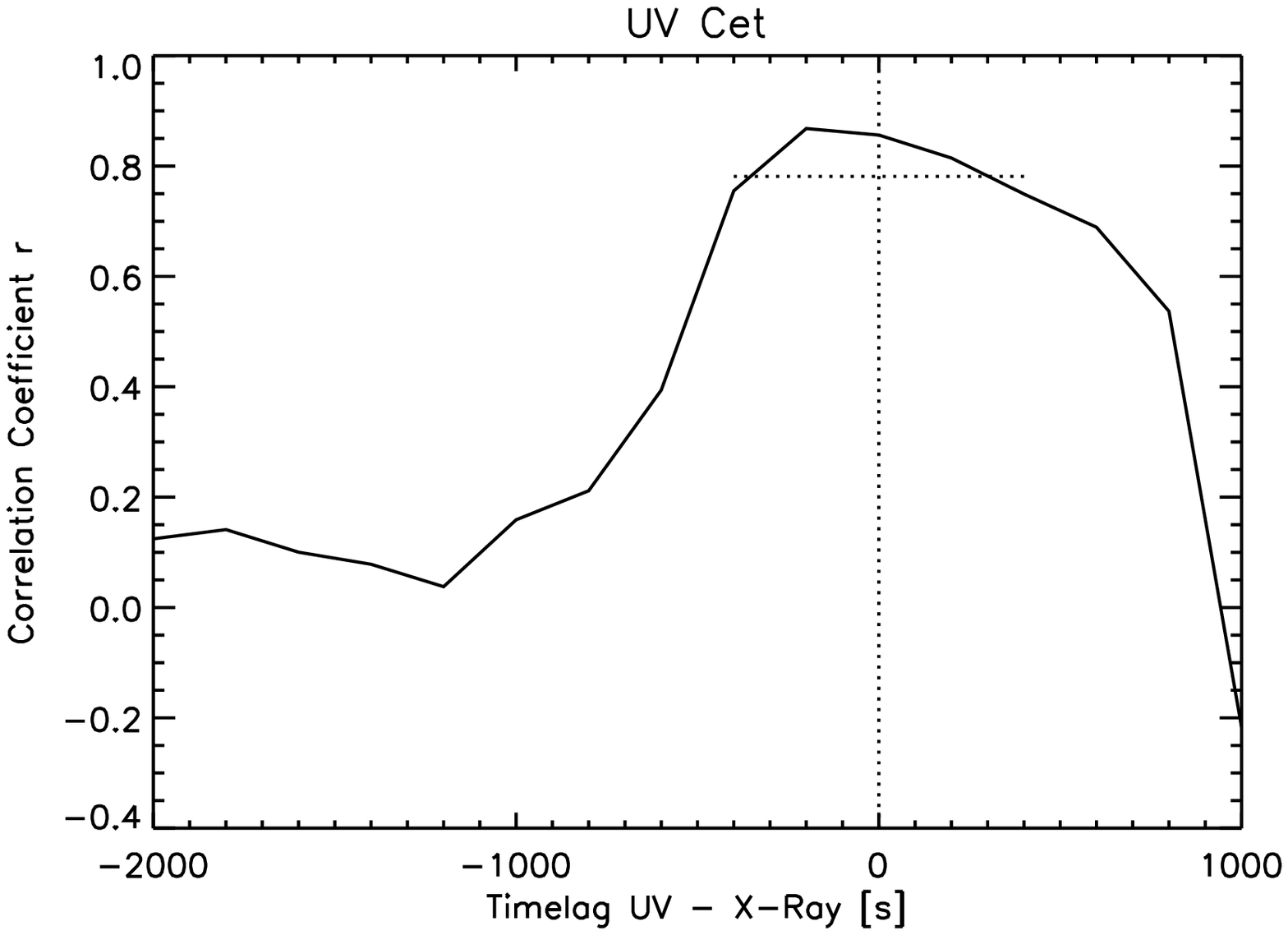,height=6.0cm,width=7.0cm}
}
}
\caption{The light curves (left panels) and their cross-correlations
  (right panels) of \object{EV Lac} and \object{UV Cet}, where the
  UVW1 filter for the OM was used. The lowest panel in each left graph
  shows the 200s-binned X-ray data, the upmost panel the UV light
  curve, and the middle panel the X-ray light curve at maximum
  correlation (shifted by the time lag) and binned to UV resolution
  (800~s integration time and a cadence of 1120~s). The count rate is
  given on the left, the luminosity on the right ordinate. The dashed
  vertical lines indicate the start and the end of the flare, the
  dotted vertical lines the time of the flare peak in the UV. The
  right panels show the cross-correlation between the X-ray and UV
  light curve for each target (see main text). The horizontal dotted
  lines are at 90\% level of each maximum and mark the peak error
  interval.} 
\label{figure1a}
\end{figure*}

%% file: 1201fig2.tex
\begin{figure*}
\centering
\vbox{
\hbox{
\psfig{figure=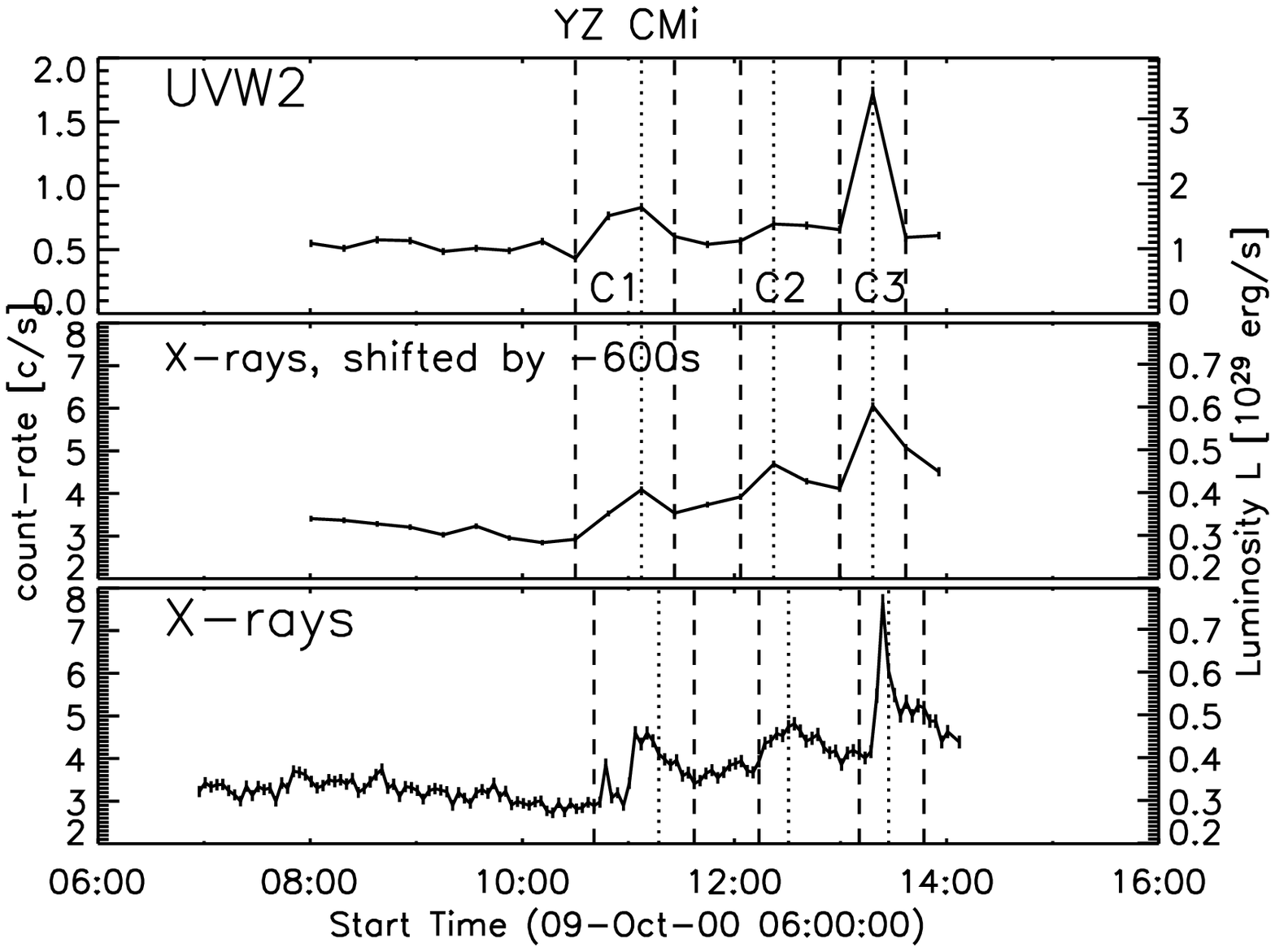,height=6.0cm,width=10.0cm}
\hspace{0.5cm}
\psfig{figure=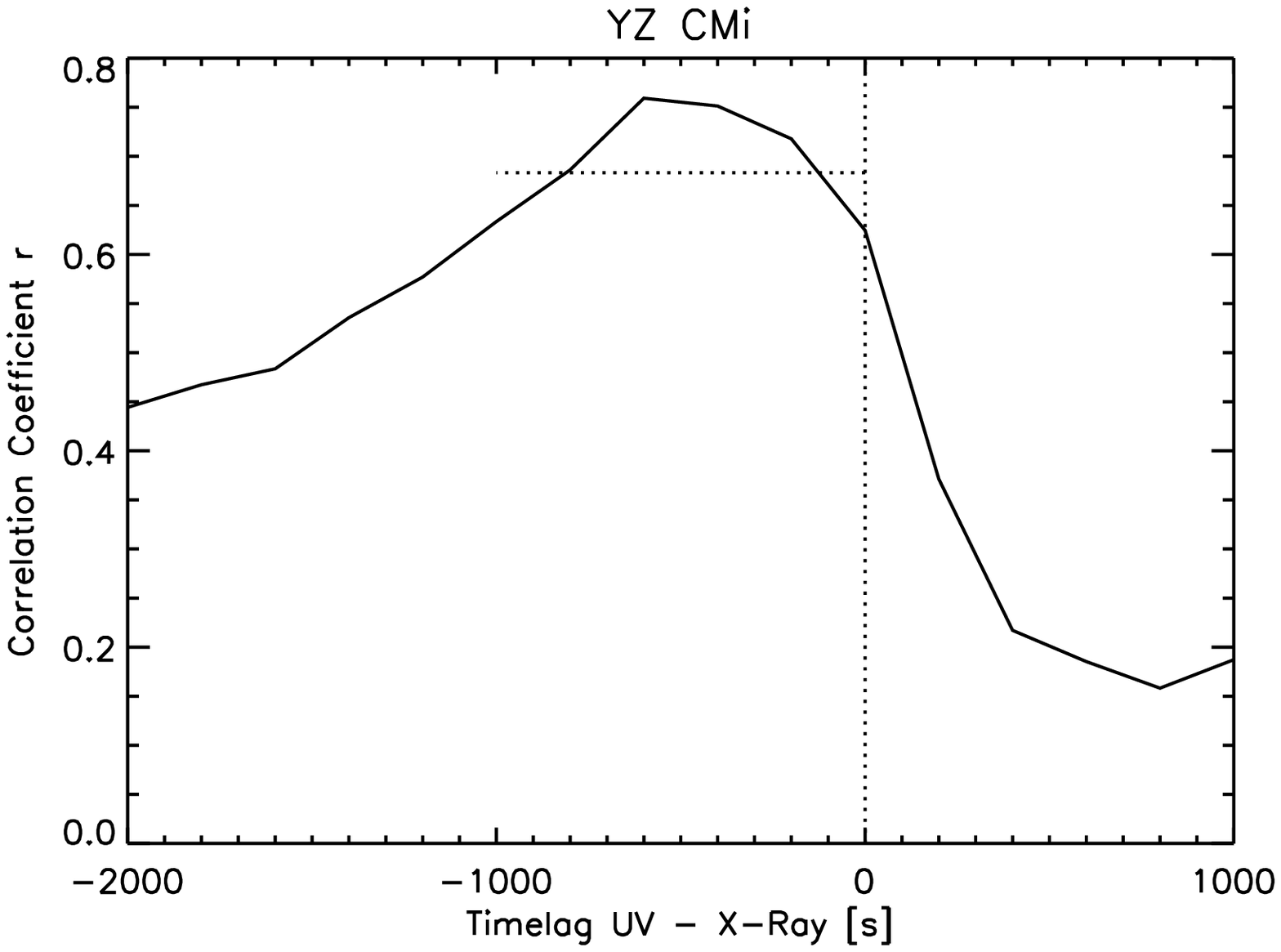,height=6cm,width=7.0cm}
}
\vspace{0.5cm}
\hbox{
\psfig{figure=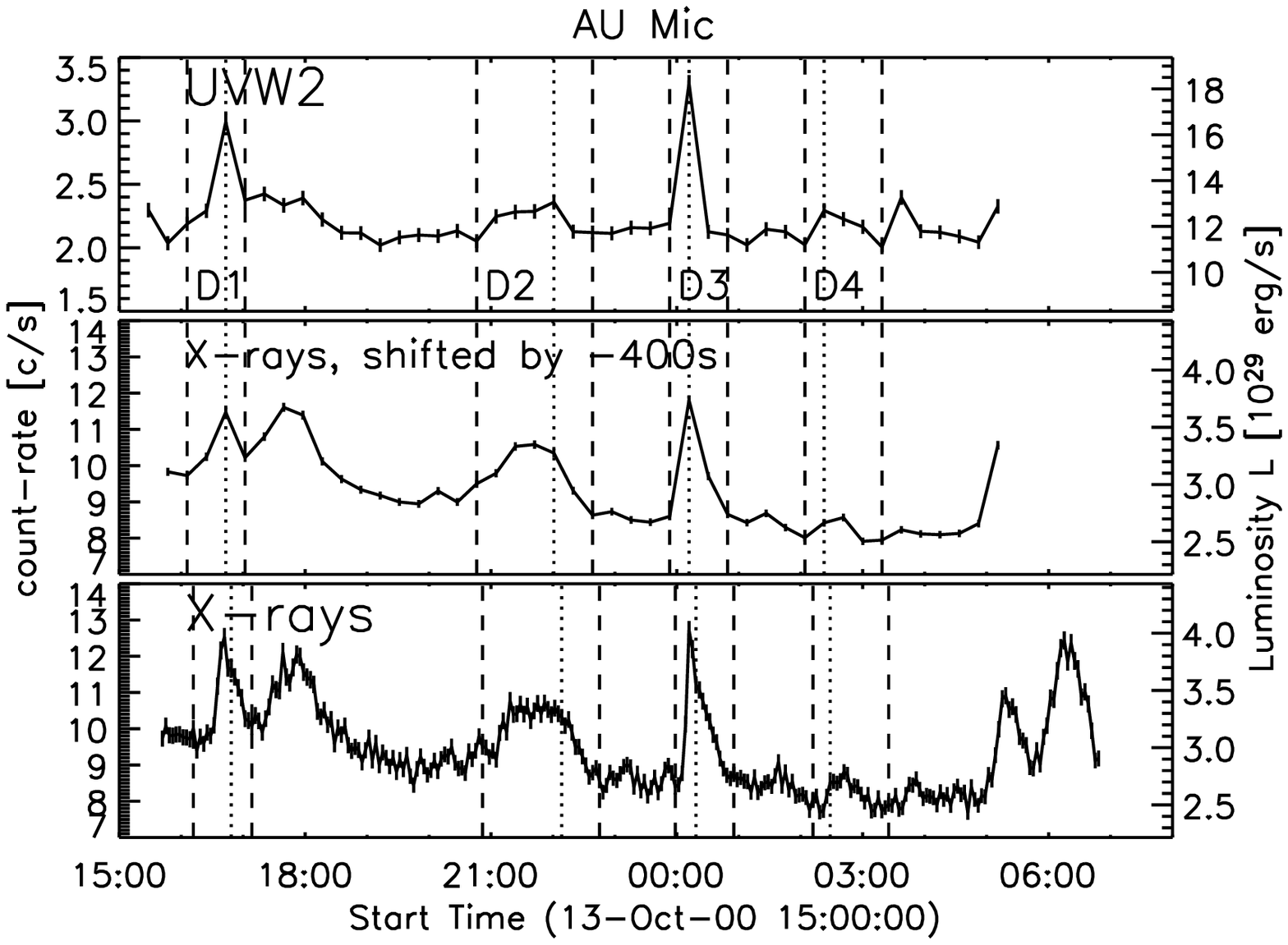,height=6.0cm,width=10.0cm}
\hspace{0.5cm}
\psfig{figure=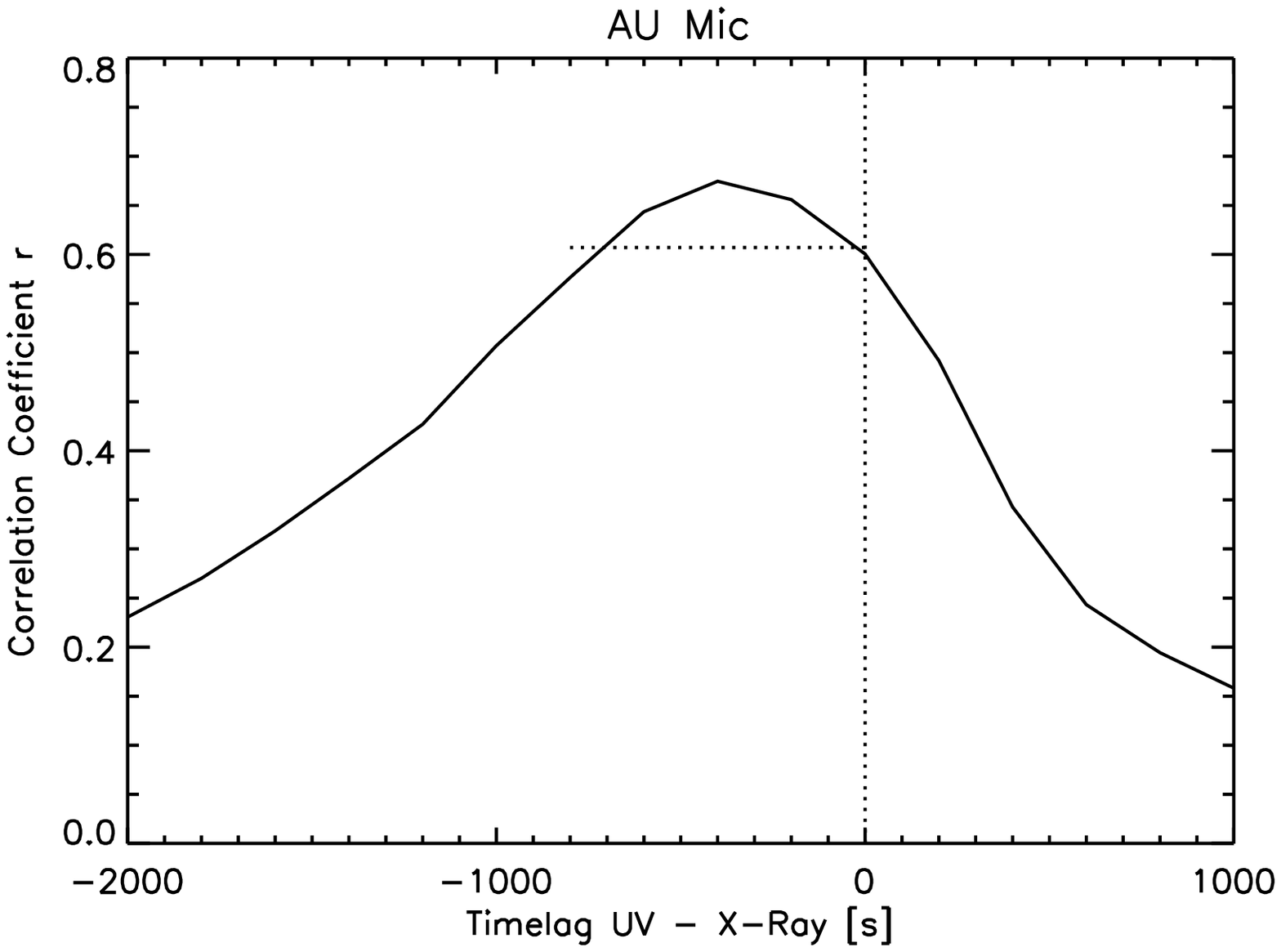,height=6.0cm,width=7.0cm}
}
\vspace{0.5cm}
\hbox{
\psfig{figure=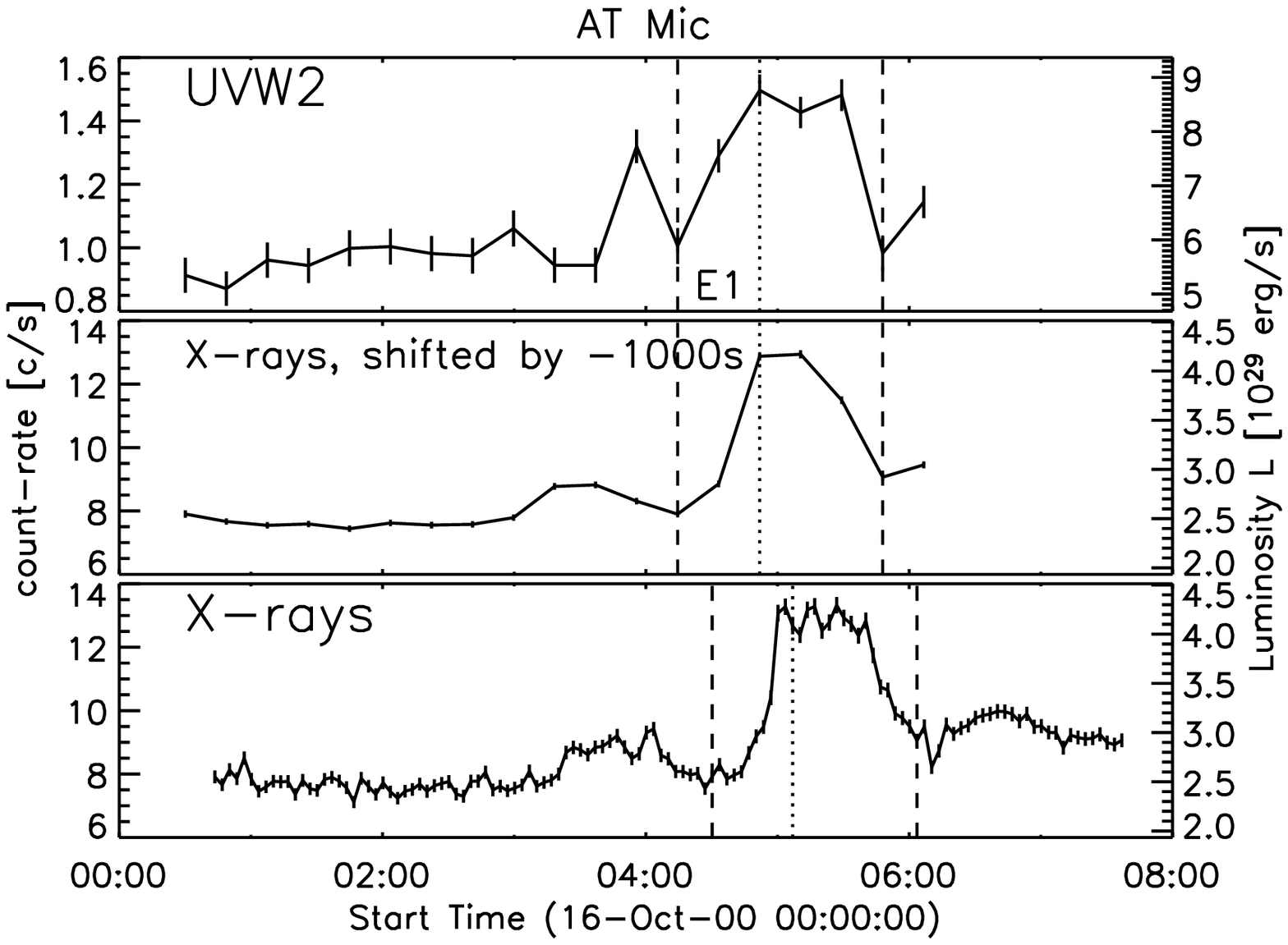,height=6.0cm,width=10.0cm}
\hspace{0.5cm}
\psfig{figure=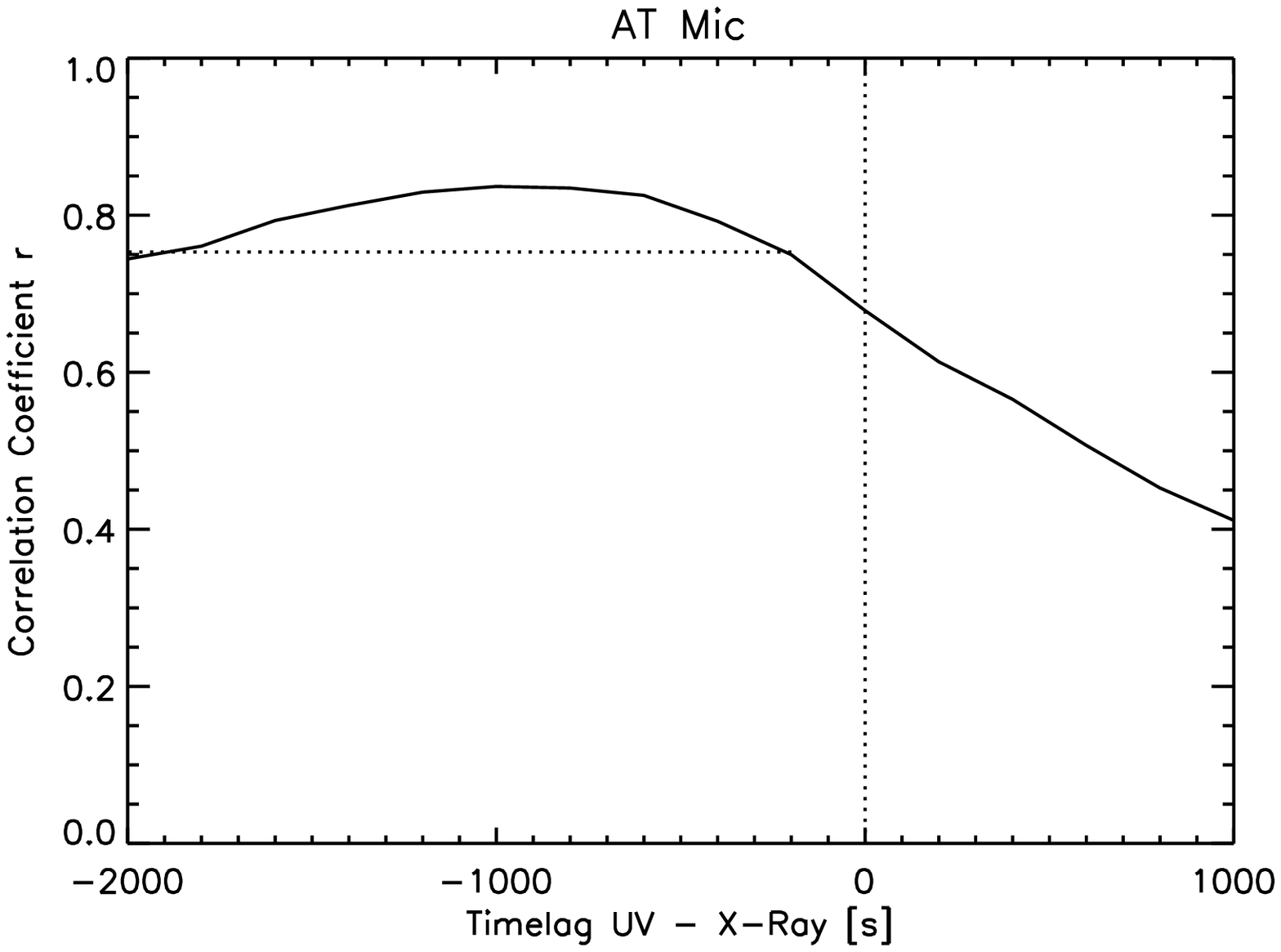,height=6.0cm,width=7.0cm}
}
}
\caption{The light curves (left panels) and their cross-correlations
  (right panels) of \object{YZ CMi}, \object{AU Mic} and \object{AT
    Mic}, where the UVW2 filter for the OM was used. The same
  explanations as in Fig.~\ref{figure1a} hold.
}
\label{figure1b}
\end{figure*}

%% file: 1201fig3.tex
\begin{figure*}[htb]
\centering
\vbox{
\hbox{
\psfig{figure=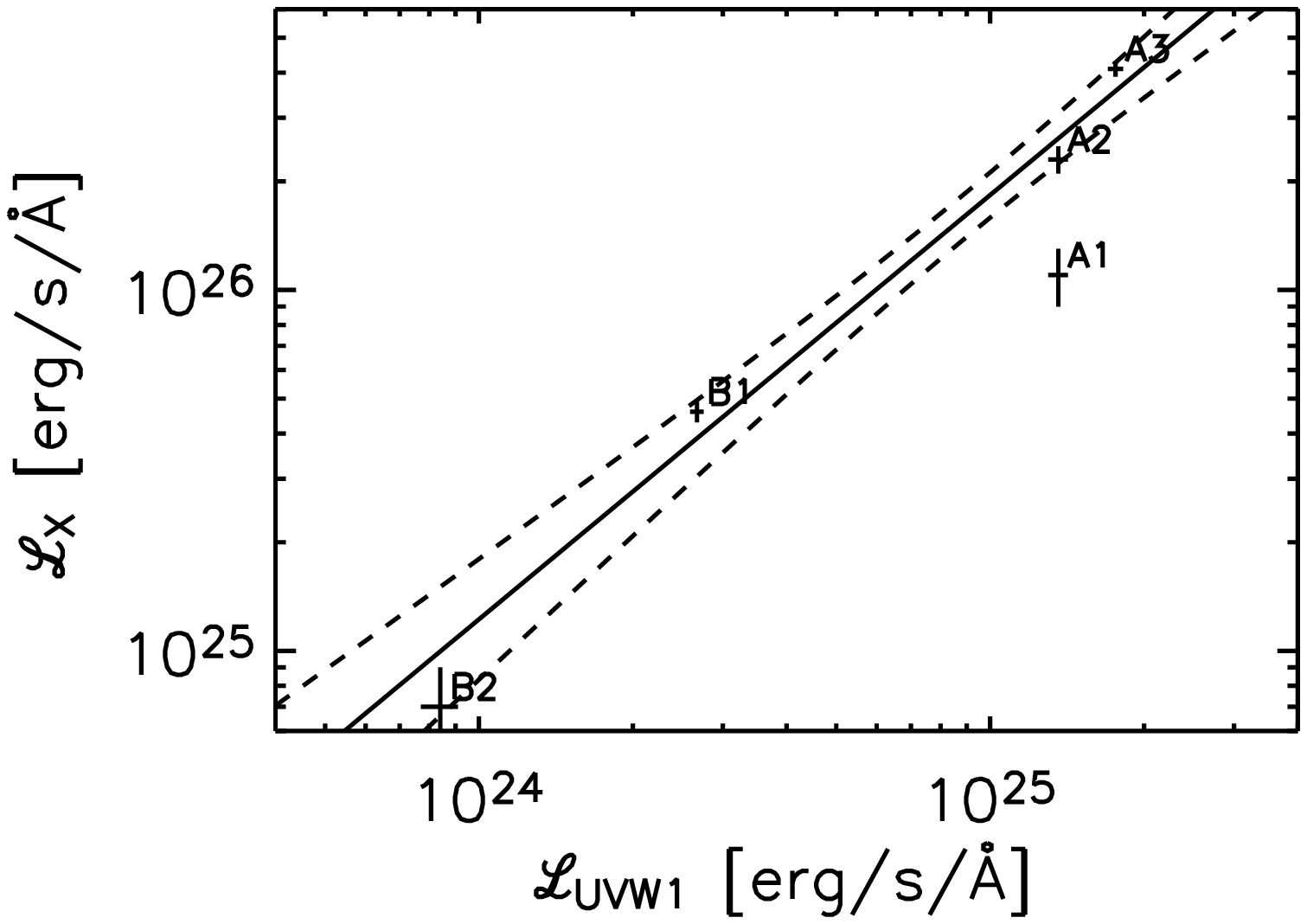,height=6.5cm,width=8.5cm}
\hspace{0cm}
\psfig{figure=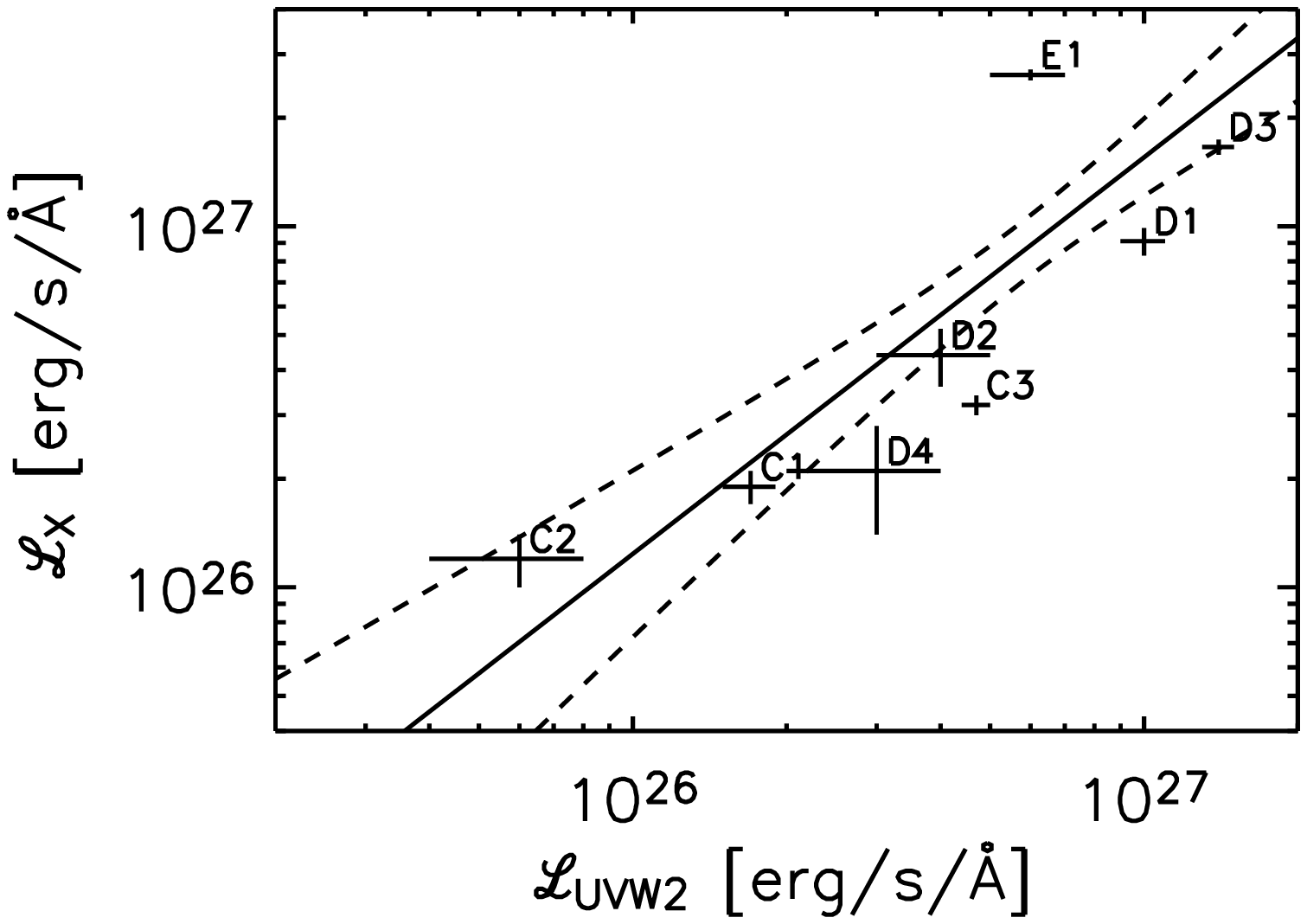,height=6.5cm,width=8.5cm}
}
\vspace{-.5cm}
\hbox{
\psfig{figure=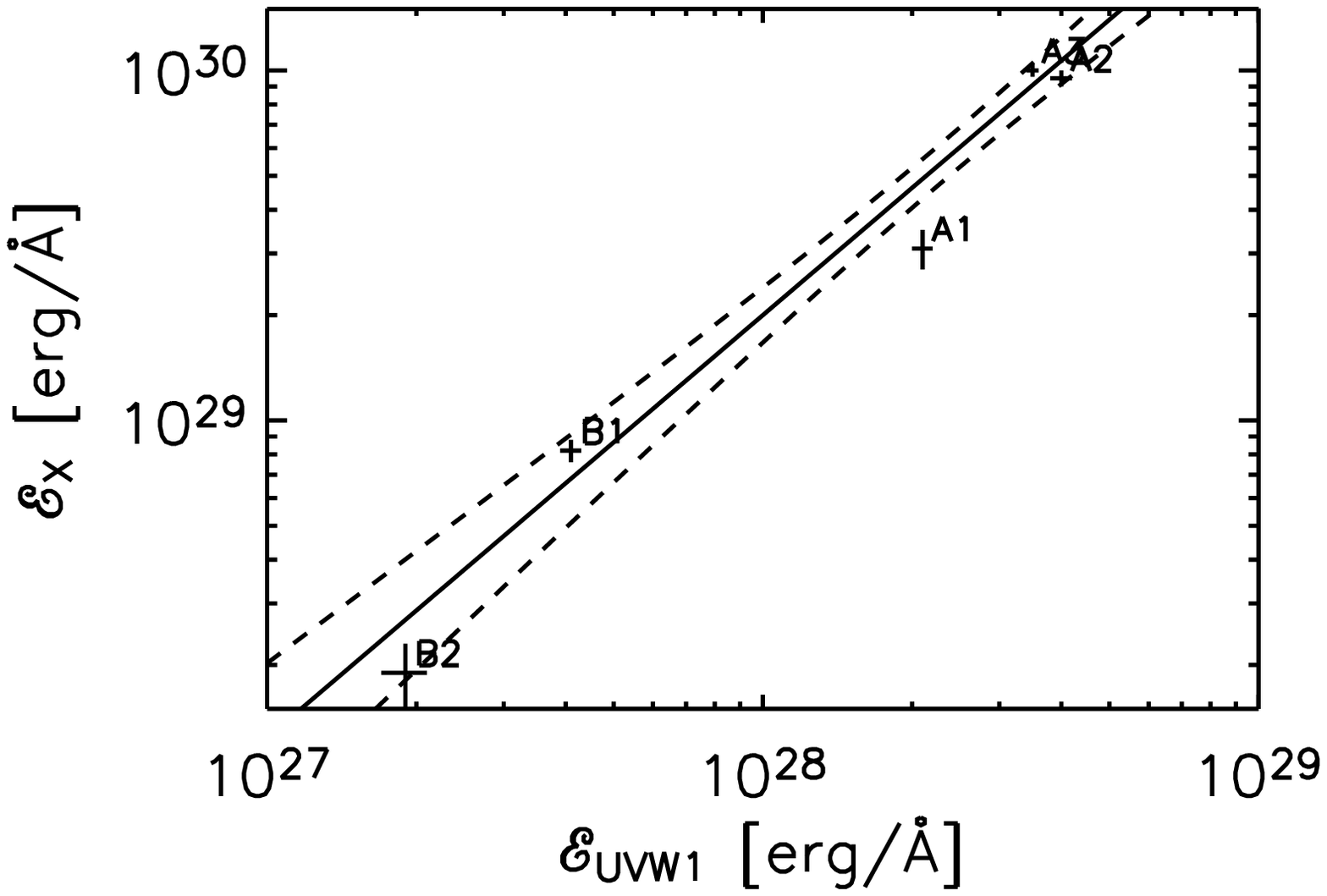,height=6.5cm,width=8.5cm}
\hspace{0cm}
\psfig{figure=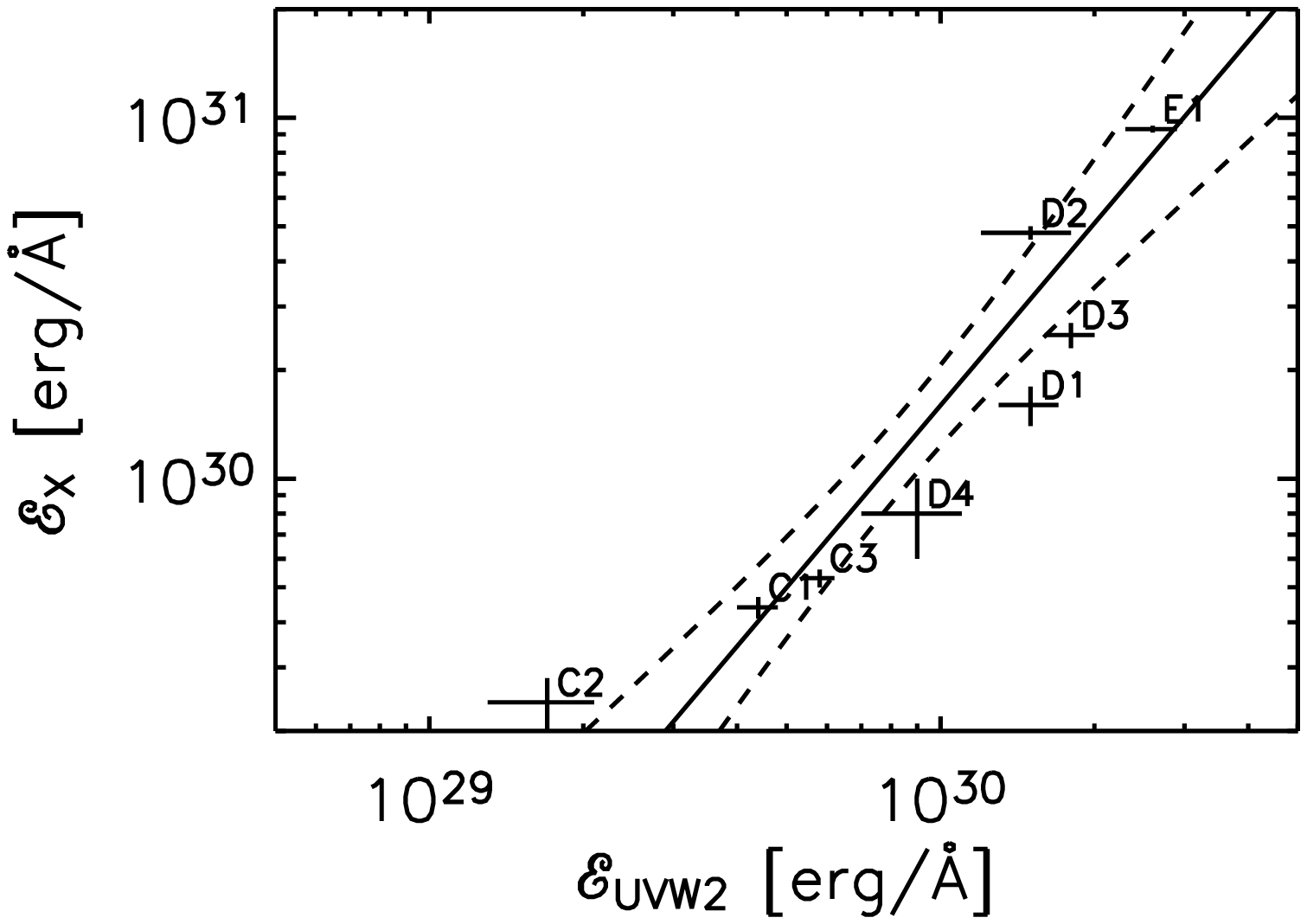,height=6.5cm,width=8.5cm}
}
}
\caption{
  X-ray vs UV flare relationships. The two panels on the left use the
  UVW1, the ones on the right the UVW2 filter. 
  The upper panels are for the luminosity increase, the lower 
  panels for the flare energy. Used throughout for the 
  regression are spectral luminosity density and spectral energy density, as
  defined in Sect.~\ref{ffr}. The crosses denote the
  flares, their sizes corresponding to the 1$\sigma$ error. They are
  labelled according to their identification as in Figs.~\ref{figure1a} and
  \ref{figure1b}. The solid line in the plot indicates the best fit
  power-law regression. The flares have been weighted with their individual
  errors for the regression. The dashed lines contour the $3\sigma$ error,
  where $\sigma$ is the standard deviation of the mean.
  The regression values are given in Table \ref{fit}.}
\label{regress_uvw1}
\end{figure*}

%% file: 1201tab2.tex
\begin{table}[t]\caption{Relationship Parameters for $\mbox
      {log(X-ray)} = \kappa \cdot {\rm log(UV)} + c$} \label{fit}
      \vspace{-2ex}
\begin{tabular}{clcc} 
\hline
\hline

& & $\kappa$ & $c$ \\
\hline

& UVW1 & 1.17$\pm$0.05 & $-$3.1$\pm$1.4 \\ 

\raisebox{2ex}[0pt]{$\mathcal{L}$} & UVW2& 1.10$\pm$0.09 &
$-$2.5$\pm$2.5  \\ 
 
\hline

& UVW1 & 1.21$\pm$0.05 & $-$4.5$\pm$1.4 \\ 

\raisebox{2ex}[0pt]{$\mathcal{E}$} & UVW2 & 1.68$\pm$0.13 &
$-$20$\pm$4 \\

\hline
\end{tabular}
\end{table}

%% file: 1201disc.tex
Comparing the peak luminosity increase as well as the total energy of
flares, we find that the UV and X-ray emissions are correlated and
that the relationships follow power laws, with power-law indices
between 1 and 2.  
For the UVW2 results, the different power-law indices for the
luminosity and energy relationships indicate that different time
scales are involved in the energy release of X-ray and UV emission.  
While the luminosity increase roughly scales with an index of unity,
the energy, which is time-integrated luminosity, scales with a
power-law index significantly larger than unity.  
This suggests that for larger flares the duration of the X-ray
emission is relatively longer than the duration of the UV emission.  
We are careful not to draw quantitative conclusions for the UVW1
results because of the extremely small statistical sample.

The Neupert effect (Eq.~(\ref{eq_neupert})) predicts a temporal
correlation between UV energy and X-ray luminosity.
Although our data do not have a high enough temporal resolution, we can
still test for a consistency with the Neupert effect indirectly. 
If the Neupert effect holds, then the total UV flare energy
should be roughly proportional to the X-ray peak luminosity increase.
From the chromospheric evaporation scenario, we indeed expect that a
larger influx of non-thermal energy results in a larger amount of hot
plasma, i.e.~in more X-ray emission. 
We test for such a correlation, which is plotted in Fig.~\ref{neupert},
and the power-law fit parameters are given in Table \ref{tab_neup}. 
\input{1201tab3} 
We find that there is such a correlation between the UV and the X-ray
emission. 
The existence of an X-ray-luminosity/UV-energy correlation is in
agreement with the Neupert-effect relationship and would suggest that
the plasma is heated from the bottom of the  magnetic flux tube to the
top, first reaching the chromosphere and only a few hundred seconds
later the hotter corona, which is consistent with chromospheric
evaporation.    
It does not, however, confirm the Neupert effect. 
For that, the time resolution would have to be much better. 
The power-law exponent is close to unity for the UVW1 flares and somewhat
larger for the UVW2 flares. 
If this correlation were indeed due to the Neupert effect, a power-law
exponent of 1 would indicate that the factor of proportionality $q$ in
Eq.~(\ref{eq_neupert}) would be similar for all flares and thus imply
similar physical conditions (e.g.~the physics of energy transport and
transformation) for all flares.  
A power-law exponent of a different value would suggest an energy
dependence of $q$.  
\input{1201fig4} 

We would like to remark here that time resolution of the light curve
also has an effect on the slope.  
While the flare energy is not strongly dependent on the bin size, the
luminosity increases are.  
If the bin size is larger than the time scale of the flare, the
underestimation of the luminosity increase can be severe.  
For X-rays, comparing the 200s-binned data with the 800s-binned data,
we find that for small flares the luminosity increase is
underestimated more than for large flares if the bins are large.  
For the UV flares, which are known to be more impulsive and have shorter
time scales than the X-ray flares, the luminosity increase might be
underestimated throughout. 
The luminosity-energy relationship is especially affected by this
underestimation and the true correlation might be less steep and
closer to unity.   

It may also be that all these correlations are just a manifestation of
the big flare syndrome (BFS), which states that statistically all
energetic flare phenomena are more intense in larger flares,
regardless of the detailed physics \citep{kahler1982}.  
On the other hand, a correlation of two a priori unrelated parameters
tells us something about the underlying physics. 
A correlation between extensive parameters (e.g.\ volume and mass) can be
regarded as trivial, however, a correlation between intensive parameters
(e.g.\ energy and luminosity) is not trivial.
Since in this paper we are studying correlations between intensive
parameters, even if the correlations are due to the BFS, they tell us
something about the underlying physics.
We can test for the BFS, if we compare the correlation coefficient of
flare parameters we expect to correlate based on flare models, like
$L_{\rm x}-L_{\rm uv}$, $E_{\rm x}-E_{\rm uv}$ and the relationship
following from a possible Neupert effect $L_{\rm x}-E_{\rm uv}$, to
flare parameters which have no physical reason to correlate ($E_{\rm
  x}-L_{\rm uv}$).  
The correlation of the latter would reflect merely the BFS.
The former being significantly higher than the latter would indicate
an additional physical cause than just the BFS. 
Indeed, for the UVW2 filter, the relationship $E_{\rm x}-L_{\rm uv}$
gives a much lower correlation coefficient ($r=0.63$) than the ones
which are physically related ($r=0.83$ for luminosity, $r=0.92$ for
energy and ``Neupert'').  
Therefore, the UVW2 correlations are probably reflecting more than
just the BFS. 
However, the correlation coefficients for the small UVW1 sample are
all above 0.95, including the BFS control parameters $E_{\rm x}-L_{\rm
  uv}$, and cannot distinguish between trivial and non-trivial
scalings.

The slopes of the average stellar luminosities are close to unity, in
agreement with \citet{mathioudakis1989}. 
If the total stellar emission in X-rays and UV is a result of the
superposition of many flares, then the X-ray to UV ratio of the average
stellar luminosity should be similar to the X-ray to UV ratios of the
flare energies.  
This is indeed so.
For the flares, the ratio of X-ray to UV spectral energy density
ranges from 10 to 29 (average of 19) for UVW1 and from 0.9 to 3.6
(average of 1.7) for UVW2.   
For the entire sample of stars, the ratio of X-ray to UV average spectral
luminosity density is  between 15 and 16 for the UVW1 filter and
between 1.6 and 3.3 for the UVW2 filter.  
The quiescent spectral luminosities have the same ratios.  
The energy ratios among the flares have a relatively wide spread because
the power-law exponent is greater than unity.
The stellar luminosity ratios within each UV wave band vary much less
and are very close to the average flare energy ratios.   
Previously, \citet{haisch1990} also reported that the ratio between
energy losses in coronal X-rays and chromospheric \ion{Mg}{ii} lines
is similar in flares and in quiescence in the dMe star \object{Proxima
  Cen}. 
This indicates a similar energy release physics both in flares and in the
low-level emission. 
A possible explanation is that the low-level emission is in fact
produced by a large number of unresolved flares that heat
chromospheric gas and build up a corona by chromospheric evaporation.  
This hypothesis has recently found strong support from statistical
light curve analysis \citep{guedel2003a,arzner2004}. 

The extremely good temporal correlation between UV and X-ray flares is
noteworthy. 
For almost every increase in UV emission, we also observe a corresponding
increase in X-rays, which is a similar result to the one found between
the optical U-band and X-rays in a recent observation of
\object{Proxima Cen} \citep{guedel2004}.   
The overall good correlation between UV and X-ray flare occurrence is
contrary to many previous observations, where some  
X-ray flares do not show a UV (or radio) counterpart and vice versa. 
This is especially true for older observations; this might be due to
poorer instrumental performance.   
On the Sun, however, soft X-ray flares (usually less intense than the
typical stellar flare observed here) only occasionally show
simultaneous white-light emission, but all white-light flares also
have a corresponding X-ray flare.  
Recent investigations by \citet{matthews2003} show that white-light
production is connected to peak pressure. 
It might be that in these more energetic stellar flare events such a
critical peak pressure is almost always present, and therefore all
flares are observed in both wavebands.  

In the UVW2 observations, we notice that there are two types of flares. 
The impulsive ones with a linear increase, an exponential decrease and
a sharp peak, and the flares with a flat top.  
The latter are rising somewhat slower than the impulsive ones, showing
prolonged, sustained peak emission. 
The tops of these flares (E1, D2, C2) are not constant, and in the case
of E1 show periodic variations.
We note that in the energy relationship plot these flat-top flares lie
above the best-fit power-law correlation, which means that either
there is a deficit in UV emission, or that there is an excess in
X-rays.   
A similar phenomenon is seen in the correlation between microwave and
hard X-ray emission in solar flares \citep{kosugi1988}, where the long
duration flares lie above the mean linear correlation, indicating a
larger microwave to hard X-ray ratio for these flares. 


%% file: 1201tab3.tex
\begin{table}[t]\caption{Neupert Relationship Parameters:
    $\mathcal{L}_{\rm x} = 10^c \cdot \mathcal{E}_{\rm uv}^\kappa$}
\label{tab_neup}  
\vspace{-2ex}
\begin{tabular}{lcc} 
\hline
\hline

& $\kappa$ & $c$ \\
\hline

UVW1 & 1.08$\pm$0.05 & $-$4.4$\pm$1.5  \\ 
 
UVW2 & 1.40$\pm$0.11 & $-$15$\pm$3  \\ 

\hline
\end{tabular}
\end{table}

%% file: 1201fig4.tex
\begin{figure}[ht]
\centering
\vbox{
\psfig{figure=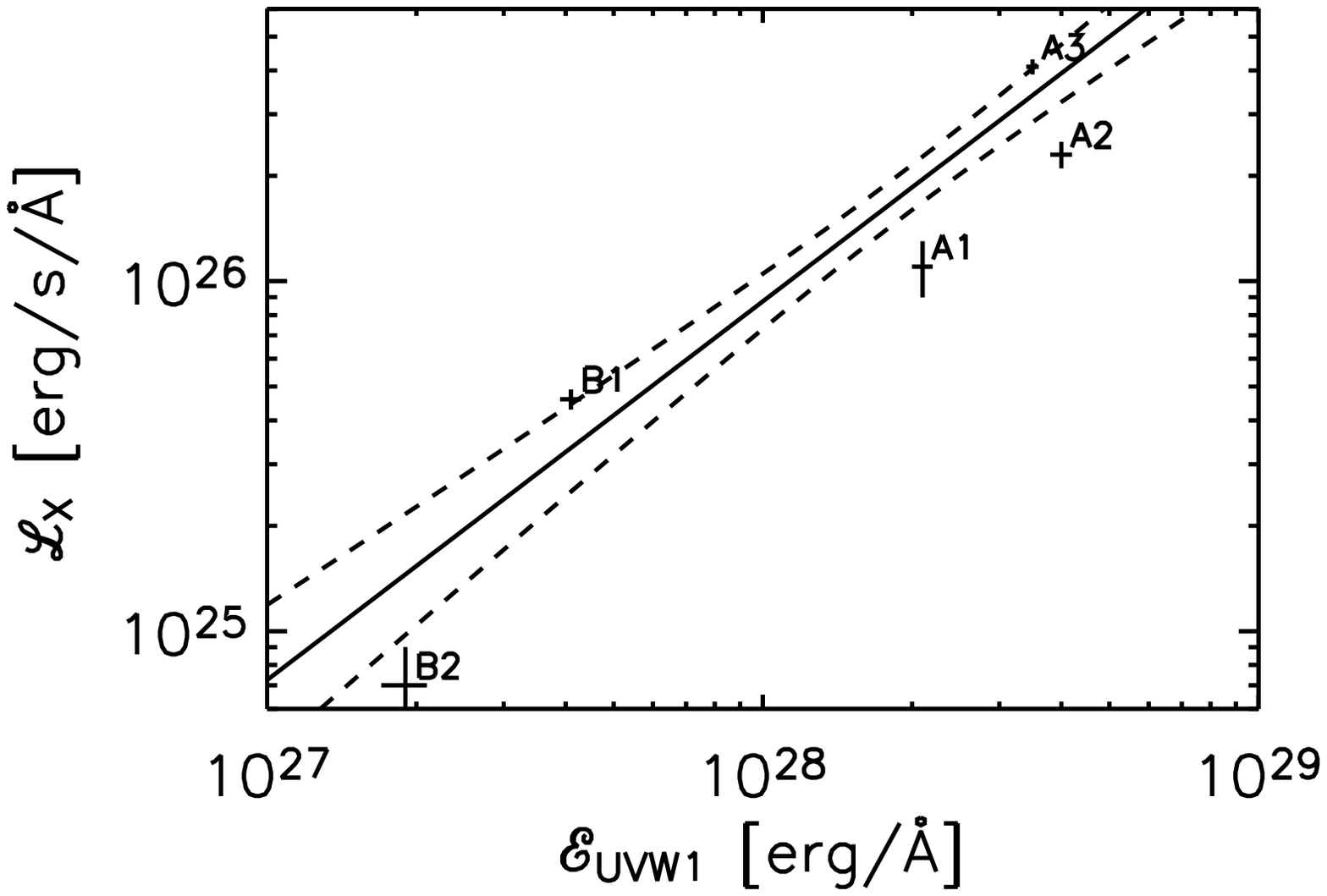,height=7cm,width=8.5cm}
\vspace{-.5cm}
\psfig{figure=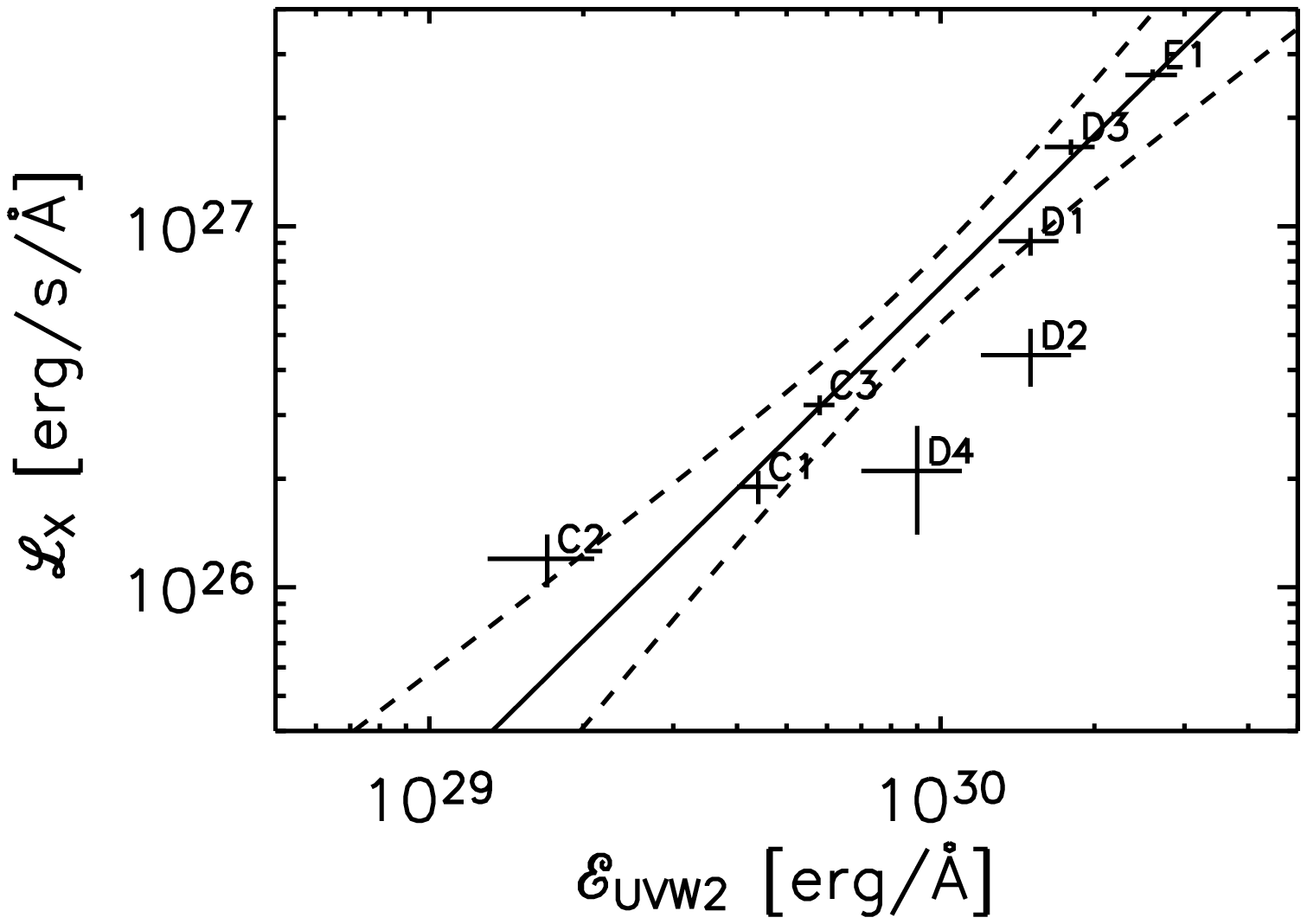,height=7cm,width=8.5cm}
}
\caption{
  Relationship between X-ray luminosity increase and UV flare
  energy. The conventions are the same as in Fig.~\ref{regress_uvw1}. 
  The fit parameters are given in Table \ref{tab_neup}.
} 
\label{neupert}
\end{figure}

%% file: 1201conc.tex
We have investigated X-ray and UV light curves of several dMe stars,
focusing on the temporal coincidence of flares and possible
correlations between fluxes and radiative energies. 
The aim of our project was to study causal relations between mechanisms
that produce UV and X-ray emission. 
The chromospheric evaporation scenario developed from solar
observations and flare simulations predicts that the impulsive-phase
optical/UV emission, most likely due to accelerated electrons
impacting on the chromosphere, should precede the more slowly evolving
soft X-ray emission that is emitted by the heated plasma.  
Specifically, the Neupert effect should approximately hold, i.e.\ the
X-ray light curve is proportional to the time integral of the
optical/UV light curve.  
Further, if the energy in the plasma stems from the accelerated
electron population, we expect that the flare amplitudes are roughly
correlated in amplitude or radiated energy. 

We find evidence of both features predicted by the evaporation scenario.
Firstly, most UV flares characteristically precede the X-ray peaks by
typically ten minutes, which approximately coincides with the soft
X-ray flare rise time.  
Secondly, we find a close near-linear correlation between the peak
fluxes of the optical and the X-ray flares. 
The correlation becomes non-linear for the total radiative energies,
implying that for larger flares the time scales of the X-ray flares
become longer compared to the UV flare time scales. 

An interesting aspect is the comparison of the X-ray to UV energy loss
rate ratio of flares with the corresponding ratio between average
emissions. 
We find that the two are indeed similar, which suggests that the total
emission is a superposition of individual flare emissions.    
Our observations thus support a picture in which stellar chromospheres
and coronae are continuously heated by impulsive energy release
processes.
